\documentclass[12pt]{article}
\usepackage{epsfig}
\usepackage[dvips]{color}
\begin{document}

\def\be{\begin{equation}}
\def\ee{\end{equation}}
\def\bq{\begin{equation}}
\def\eq{\end{equation}}
\def\bqa{\begin{eqnarray}}
\def\eqa{\end{eqnarray}}
\def\roughly#1{\mathrel{\raise.3ex
\hbox{$#1$\kern-.75em\lower1ex\hbox{$\sim$}}}}
\def\lsim{\roughly<}
\def\gsim{\roughly>}
\def\llgm{\left\lgroup\matrix}
\def\rrgm{\right\rgroup}
\def\vectrl #1{\buildrel\leftrightarrow \over #1}
\def\partrl{\vectrl{\partial}}
\def\gslash#1{\slash\hspace*{-0.20cm}#1}



\begin{center}
{\LARGE\bf
Unique Prediction for\\ High-Energy {\boldmath$J/\psi$} Photoproduction:\\
\vspace*{0.3cm}
Color Transparency and Saturation.}
\end{center}
 \vspace {0.5 cm}
\begin{center}
{\bf Dieter Schildknecht} \\[2.5mm]
Fakult\"{a}t f\"{u}r Physik, Universit\"{a}t Bielefeld \\[1.2mm] 
D-33501 Bielefeld, Germany\\[1.2mm] 
and \\[1.2mm] 
Max-Planck-Institut f\"ur Physik (Werner-Heisenberg-Institut),\\[1.2mm] 
F\"ohringer Ring 6, D-80805 M\"unchen, Germany

\end{center}

\vspace{2 cm}

\baselineskip 20pt

\begin{center}
{\bf Abstract}
\end{center}
\begin{quote}
Based on the color-dipole picture, we present a successful parameter-free
prediction for the recent high-energy $J/\psi$ photoproduction data from
the Large Hadron Collider. The experimental data provide 
empirical evidence for the
transition from color transparency to saturation.
\end{quote}
\bigskip

Inspired by experimental data \cite{LHCb} on $J/\psi$ photoproduction, 
$\gamma p \to J/\psi~p$, at
very high energies obtained at the Large Hadron Collider (LHC), we
revive our predictions \cite{PLB638, EPJC37} on $J/\psi$ production based on the
color-dipole picture (CDP)\footnote{Compare ref. \cite{PRD85} and the list
of literature on the CDP quoted there.}.
 As depicted in figs. 1 and 2,
the total photoabsorption cross section and the forward production of
quark-antiquark states, $\gamma^* p \to (q \bar q)^{J=1}_{T,L} p$,
of total spin $J=1$ at low
values of $x \cong Q^2/W^2 \ll 1$ are represented by \cite{EPJC37,PRD85}
\begin{eqnarray}
& & \sigma_{\gamma^*_{T,L}p} (W^2 , Q^2) = \label{1} \\
& & = \int dz \int d^2 r_\bot 
| \psi_{T,L} (r_\bot , z(1-z), Q^2) |^2 
\sigma_{(q \bar q)^{J=1}_{T,L}p}
(\vec r_\bot \sqrt{z(1-z)}, W^2) 
\nonumber
\end{eqnarray} 
and
\begin{eqnarray}
& & \frac{d\sigma_{\gamma^*_{T,L}p \to (q \bar q)^{J=1}_{T,L}p}}{dt}
(W^2,Q^2) \bigg|_{t=0} 
= \label{2} \\
& & = \frac{1}{16 \pi} \int dz \int d^2 r_\bot \left| \psi_{T,L}
(r_\bot , z (1-z), Q^2) \right|^2
\sigma^2_{(q \bar q)^{J=1}_{T,L} p}
(\vec r_\bot \sqrt{z (1-z)}, W^2).
\nonumber
\end{eqnarray}
\begin{figure}[h]
\begin{center}
\epsfig{file=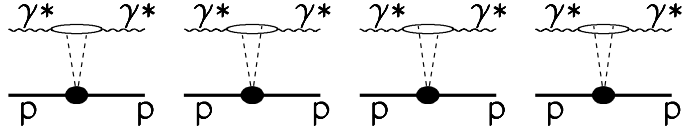, width=12cm}
\caption{The forward Compton amplitude}
\end{center}
\end{figure}
\begin{figure}[h]
\begin{center}
\epsfig{file=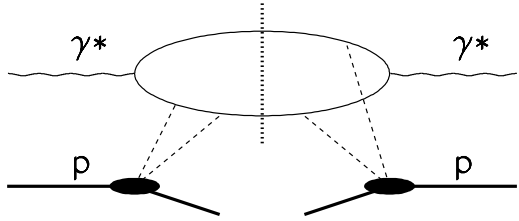, width=12cm}
\caption{One of the 16 diagrams for diffractive production. The vertical line
indicates the unitarity cut corresponding to the diffractively produced
final states, $(q \bar q)^J$. Production of (discrete or continuum) vector
states corresponds to $(q \bar q)^J$ production with $J=1$.}
\end{center}
\end{figure}
In standard notation, $\vert\psi_{T,L} (r_\bot, z(1-z),Q^2)\vert^2$
denotes the probability for the photon of virtuality $Q^2$ to couple
to a $(q \bar q)^{J=1}_{T,L}$ state specified by the transverse size
$\vec r_\bot$ and the longitudinal momentum partition $0\le z \le 1$,
and $\sigma_{(q \bar q)^{J=1}_{T,L}p}  \left( \vec r_\bot 
\sqrt{z(1-z)}, W^2 \right)$ denotes the $(q \bar q)$-dipole-proton cross
section at the total $\gamma^*p$ center-of-mass energy $W$. The
two-gluon coupling of the $q \bar q$ dipole state requires a representation
of the $(q \bar q)$-color-dipole cross section of the form 
\cite{EPJC37,PRD85}
\be
\sigma_{(q \bar q)^{J=1}_{T,L}p} (\vec r_\bot \sqrt{z(1-z)}, W^2) = \int d^2 l^
\prime_\bot \bar\sigma_{(q \bar q)^{J=1}_{T,L}p} (\vec l^{~\prime 2}_\bot , 
W^2)(1 - e^{-\vec l^{~\prime}_\bot \cdot \vec r_\bot \sqrt{z(1-z)}}) .
\label{3}
\ee
Upon introducing the mass, $M$, of the $(q \bar q)^{J=1}$ states, and 
upon applying the optical theorem to relate $(q \bar q)^{J=1} p$ forward
scattering to the $(q \bar q)^{J=1} p$ total cross section, we identically
reformulate the total photoabsorption cross section (\ref{1}) in terms
of the $(q \bar q)p$ forward-production cross section (\ref{2}), 
obtaining \cite{PRD66}\footnote{A correction related to the contributions
due to a (small) real part of the $q \bar q$ forward-production amplitude
is ignored in (\ref{4}) and (\ref{5}).}
\be
\sigma_{\gamma^*_T p} (W^2 , Q^2) = \sqrt{16\pi} 
\sqrt{\frac{\alpha R_{e^+ e^-}}{3\pi}} \int_{m^2_0} dM^2 \frac{M}{Q^2 + M^2} 
\sqrt{\left.\frac{d\sigma_{\gamma^*_T p
\rightarrow (q \bar q)^{J=1}_T p}}{dt dM^2}\right|_{t=0}} 
\label{4} 
\ee
and
\be
\sigma_{\gamma^*_L p} (W^2 , Q^2) = \sqrt{16\pi} 
\sqrt{\frac{\alpha R_{e^+ e^-}}{3\pi}} \int_{m^2_0} dM^2 \frac{\sqrt{Q^2}}
{Q^2 + M^2} 
\sqrt{\left.\frac{d\sigma_{\gamma^*_L p
\rightarrow (q \bar q)^{J=1}_L p}}{dt dM^2}\right|_{t=0}} .
\label{5} 
\ee
The representations (\ref{4}) and (\ref{5}) explicitly relate the
transverse and longitudinal total photoabsorption cross sections to
the diffractive forward-production cross sections,
\be
\frac{d \sigma_{\gamma^*_{T,L}p \to (q \bar q)^{J=1}_{T,L}p}}{dtdM^2}
\bigg|_{t = 0}  = 
\frac{d \sigma_{\gamma^*_{T,L}p \to (q \bar q)^{J=1}_{T,L}p}}{dtdM^2}
(W^2, Q^2, M^2)  \bigg|_{t = 0}  ,
\label{6}
\ee
of a continuum of $(q \bar q)^{J=1}$ states of mass $M$. The lower
limit $m^2_0$ in (\ref{4}) and (\ref{5}), via quark-hadron 
duality\footnote{Massless quarks are used in the derivation of (\ref{4})
and (\ref{5}).}, for
the (dominant) contribution due to light quarks fulfills $m_0 < m_\rho$,
where $m_\rho$ is the $\rho$-meson mass, and $R_{e^+e^-} = 3 \sum_q Q^2_q$,
where the sum runs over the actively contributing quark flavors, and
$Q_q$ denotes the quark charge.

The representations (\ref{4}) and (\ref{5}) for the photoabsorption cross
section explicitly demonstrate that low-x photoabsorption is due to
the coupling of the photon to $(q \bar q)^{J=1}$ (vector) dipole states
that subsequently propagate and interact via coupling to two gluons with
the gluon field in the nucleon. The massive $q \bar q$ continuum
contains the low-lying vector mesons, $\rho^0, \omega, \phi, J/\psi, \Upsilon$
etc. followed by more massive $J=1$ true continuum states, 
$X^{J=1}_{q \bar q}$, also
diffractively produced via $\gamma^*p \to X^{J=1}_{(q \bar q)}p$. For an
appropriate magnitude of the kinematic variables, the $X^{J=1}_{(q \bar q)}$
states appear as two-jet states with angular distribution characteristic
for spin $J = 1$.

For quantitative predictions, $\bar \sigma_{(q \bar q)^{J=1}_{T,L}p} (\vec
l^{~\prime 2}_\bot, W^2)$ in (\ref{3})
must be specified. Agreement with experiment for the total 
photoabsorption cross section in (\ref{1}), and, equivalently, in 
(\ref{4}) and (\ref{5}), then determines the diffractive
(forward) production of vector states and allows for a parameter-free
prediction of their photoproduction cross section (\ref{6}). Indeed,
for vector-meson production, specifically for $J/\psi$ 
production\footnote{The generalization to $\Upsilon$-production is straight
forward}, we have
\cite{PLB638, EPJC37} 
\bqa
& & \frac{d \sigma_{\gamma^*p \to J/\psi~ p}}{dt} (W^2, Q^2) \left|_{t=0} =
\right. \nonumber \\
& & \int_{\Delta M^2_{J/\psi}} dM^2 \int^{z_+}_{z_-} dz 
\frac{d \sigma_{\gamma^*p \to (c \bar c)^{J=1}p}}{dt~ dM^2~ dz} (W^2, Q^2, z, 
m^2_c, M^2),
\label{7}
\eqa
where
\be
z_\pm = \frac{1}{2} \pm \frac{1}{2} 
\sqrt{1-4 \frac{m^2_c}{M^2}}.
\label{8}
\ee
In (\ref{7}), $m_c$ denotes the mass of the charm quark, and
$M^2 \equiv M^2_{c \bar c}$ denotes the mass squared of the 
$c \bar c$ state in the interval $\Delta M^2_{J/\psi}$ determined
by the level spacing of the $c \bar c$ bound-state spectrum of states
observed in $e^+e^-$ annihilation.

Replacing the cross section for the open-charm continuum on the 
right-hand side in (\ref{7}) by the cross section  for
$c \bar c$ production at threshold, 
\bqa
\frac{d \sigma_{\gamma^* p \to (c \bar c)^{J=1}p}}{dt~dM^2 dz}
\hspace*{-0.5cm} && (W^2, Q^2, z, m^2_c, M^2) \nonumber \\
&& \longrightarrow \frac{d\sigma}{dt~dM^2~dz} \left(W^2, Q^2, z = \frac{1}{2},
M^2 = 4m^2_c = M^2_{J/\psi} \right),
\label{9}
\eqa
the integration in (\ref{7}) simplifies to become \cite{EPJC37}
\be
\int^{4m^2_c+\Delta M^2_{J/\psi}}_{4 m^2_c} dM^2 \int^{z_+}_{z_-} dz =
\int^{4m^2_c+\Delta M^2_{J/\psi}}_{4 m^2_c} dM^2 \sqrt{1 - \frac{4m^2_c}{M^2}}
\equiv \Delta F^2(m^2_c, \Delta M^2_{J/\psi}),
\label{10}
\ee
where
\begin{eqnarray}
 \Delta F^2(m_c^2,\Delta M_{J\psi}^2)&\equiv & \int^{4m^2_c + \Delta M^2_{J/\psi}}_{4m^2_c} \, 
     d M^2 \sqrt{1-\frac{4m^2_c}{M^2}} \int^1_0 \, dy  \label{11} \\
  &=& (4m^2_c + \Delta M^2_{J/\psi}) \sqrt{\frac{\Delta M^2_{J/\psi}}
     {4m^2_c + \Delta M^2_{J/\psi}}} + 2 m^2_c \, 
      \ln \frac{1 - \sqrt{\frac{\Delta M^2_{J/\psi}}{4m^2_c + \Delta 
      M^2_{J/\psi}}}}{1 + \sqrt{\frac{\Delta M^2_{J/\psi}}{4m^2_c + \Delta M^2_{J/\psi}}}}  .
\nonumber
\end{eqnarray}
With (\ref{9}) to (\ref{11}), the cross section (\ref{7}) becomes
\bqa
\frac{d \sigma_{\gamma^* p \to J/\psi p}}{dt}
\hspace*{-0.5cm} && (W^2, Q^2) \bigg|_{t=0}  \label{12} \\
&& \longrightarrow \frac{d\sigma}{dt~dM^2~dz} \left(W^2, Q^2, z = \frac{1}{2},
M^2 = 4m^2_c = M^2_{J/\psi} \right) \Delta F^2 (m^2_c, \Delta M^2_{J/\psi}).
\nonumber
\eqa
The cross section for $J/\psi$ electroproduction in (\ref{12}) is
accordingly represented by charm-quark pair production at threshold
multiplied by the factor $\Delta F^2 (m^2_c, \Delta M^2_{J/\psi})$
of dimension $\Delta M^2_{J/\psi}$.

In refs. \cite{PLB638},\cite{EPJC37} we have evaluated the 
$J/\psi$-production cross
section (\ref{12}) upon specifying the dipole cross section in (\ref{3})
by the ansatz
\be 
\bar\sigma_{(q \bar q)^{J=1}_T p} (\vec l^{~\prime 2}_\bot , W^2) = 
\bar\sigma_{(q \bar q)^{J=1}_L p} (
\vec l^{~\prime 2}_\bot , W^2) = \sigma^{(\infty)} 
(W^2) \frac{1}{\pi} \delta
(\vec l^{~\prime 2}_\bot - \Lambda^2_{sat} (W^2)) , 
\label{13}
\ee 
previously employed \cite{PRD85} in our (successful) representation 
of the body of
experimental data on the total
photoabsorption cross section (\ref{1}), or equivalently,
(\ref{4}) and (\ref{5}). The (asymptotic) dipole cross section
$\sigma^{(\infty)} (W^2)$ depends logarithmically on $W^2$.
The specification of $\sigma^{(\infty)} (W^2)$ and of the ``saturation scale''
$\Lambda^2_{sat} (W^2)$ will be given below.

With (\ref{13}), one finds that the cross section (\ref{12}) takes the
simple explicit form \cite{PLB638}
\bqa
&&\frac{d \sigma_{\gamma^*p \to J/\psi~p}}{dt} (W^2, Q^2) \left|_{t=t_{{\rm
        min}}\cong 0} 
\right. = 
\frac{3}{2} \frac{1}{16 \pi} \frac{\alpha~R^{(J/\psi)}}{3 \pi} 
\left(\sigma^{(\infty)} (W^2)\right)^2  \nonumber \\
&& \times  \frac{\Lambda^4_{sat}(W^2)}{(Q^2 + M^2_{J/\psi})^3}
\frac{1}{\left( 1 + \frac{\Lambda^2_{sat} (W^2)}{Q^2 + M^2_{J/\psi}}\right)^2}
\Delta F^2 (m^2_c, \Delta M^2_{J/\psi}),
\label{14}
\eqa
where $R^{(J/\psi)} = \frac{4}{3}$.

We stress again that (\ref{14}) provides a parameter-free prediction for
$J/\psi$ photo- and electroproduction, once the dipole cross section,
$\sigma^{(\infty)} (W^2)$, and the saturation scale, $\Lambda^2_{sat}
(W^2)$, in (\ref{13}) are specified from agreement with experiment of
the CDP prediction for the total photoabsorption cross section 
$\sigma_{\gamma^*p} (W^2,Q^2)$ in (\ref{1}). The total $J/\psi$ production
cross section is obtained from (\ref{14}) by multiplication \cite{PLB638}
with the inverse of the experimentally known slope parameter of the $J/\psi$
production diffraction peak.

In refs. \cite{PLB638, EPJC37} we compared the threshold-approximation 
result (\ref{14}) with the (numerical) results obtained without relying
on the threshold approximation (\ref{9}). We found that (\ref{14}), upon
multiplication by an overall factor of magnitude $2/3$, corresponding to
$3/2 \to 1$ in (\ref{14}), is in good agreement with the results that
do not rely on the threshold approximation (\ref{9}). In addition, in ref.
\cite{PLB638}, the cross section in (\ref{14}) was modified by the 
replacement of $\sigma^{(\infty)} (W^2)$ by $\sigma^{\prime (\infty)} (W^2)
= const~~ \sigma^{(\infty)} (W^2)$, where the constant factor takes care
of a so-called ``skewing correction'' and a correction for a small real
part in the $J/\psi$ production amplitude. The agreement with the ZEUS
experimental data \cite{ZEUS} for the $Q^2$ and the $W$ dependence of
$J/\psi$ production was elaborated upon in detail in ref. \cite{PLB638}.

In the present note, we concentrate on photoproduction, $Q^2 = 0$, and
in particular we concentrate on the experimental results on the high-energy
dependence of $\sigma_{\gamma p \to J/\psi p} (W)$, for $W \gsim 100 GeV$,
recently presented by the LHCb collaboration \cite{LHCb}. The
experimental data are reproduced in fig. 3,
\begin{figure}
\begin{center}
\epsfig{file=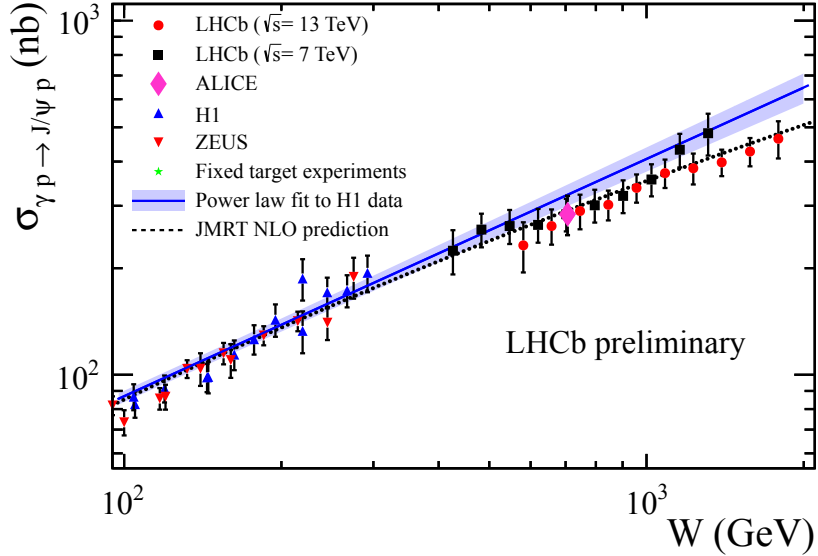,width=12cm}
\end{center}
\caption{The experimental results \cite{LHCb} from the LHCb collaboration. 
The figure is taken from ref. 1. The theoretical results from the CDP
(compare (\ref{14}) as well as (\ref{15})) given in the last column
of Table 1, are seen to agree with the experimental results from the
LHCb collaboration.}
\end{figure}
taken from ref. \cite{LHCb}. At the energy of $W = W_1 = 100 GeV$, the
LHCb result of $\sigma_{\gamma p \to J/\psi p} (W = W_1 = 100 GeV) \cong
80 nb$ is consistent with our theoretical prediction \cite{PLB638} and
with the ZEUS experimental results \cite{ZEUS}. With respect to the
LHCb results in fig. 3, we can accordingly restrict ourselves to an
analysis of the energy variation of the measured cross section relative
to the cross section at $W = W_1 = 100 GeV$. From (\ref{14}), at $Q^2 = 0$, 
for $\sigma_{\gamma p \to J/\psi p} (W^2)$ in terms of $\sigma_{\gamma p
\to J/\psi p} (W^2_1)$, we find
\bqa
\sigma_{\gamma p \to J/\psi p} (W^2) & = & \frac{\left(1+
  \frac{M^2_{J/\psi}}{\Lambda^2_{sat} (W^2_1)}\right)^2}{\left(1+
\frac{M^2_{J/\psi}}{\Lambda^2_{sat} (W^2)} \right)^2}
\frac{\left(\sigma^{(\infty)} (W^2)\right)^2}{\left(\sigma^{(\infty)}
(W^2_1) \right)^2} \sigma_{\gamma p \to J/\psi p} (W^2_1 = (100 GeV)^2),
\nonumber \\
& \equiv & F_A (\Lambda^2_{sat} (W^2)) F_B (W^2) \sigma_{\gamma p \to
J/\psi p} (W^2_1 = (100 GeV)^2),
\label{15}
\eqa
where $\sigma_{\gamma p \to J/\psi p} (W^2_1 = (100 GeV)^2) = 80 nb$ is
to be inserted on the right-hand side in (\ref{15}).

The dominant energy dependence of $J/\psi$ photoproduction is due to
the first factor, $F_A (\Lambda^2_{sat} (W^2))$, on the right-hand side
in (\ref{15}).
Inserting $M_{J/\psi} = 3.1 GeV$, and \cite{PLB638}
\be
\Lambda^2_{sat} (W^2) = B^\prime \left( \frac{W^2}{1 GeV^2} \right)^{C_2},
\label{16}
\ee
where
\bqa
B^\prime & = & 0.340 GeV^2, \nonumber \\
C_2 & = & 0.276,
\label{17}
\eqa
we find the results for $F_A (\Lambda^2_{sat} (W^2))$ shown in the fourth
column of Table. 1.
\begin{table}
\begin{tabular}{|r|l|l|l|l|l|}
\hline
$W [GeV]$ & $\Lambda^2_{sat} (W^2) [GeV^2]$ 
& $\frac{M^2_{J/\psi}}{\Lambda^2_{sat} (W^2)}$
& $F_A (\Lambda^2_{sat} (W^2))$ 
& $F_B (W^2)$ 
& $\sigma_{\gamma p \to J/\psi p} (W) [nb]$  \\
\hline
100 & 4.32 & 2.22 & 1 & 1 & 80  \\
300 & 7.92 & 1.21 & 2.12 & 1.02 & 173 \\
1000 & 15.4 & 0.624 & 3.93 & 1.11 & 349 \\
2000 & 22.6 & 0.425 & 5.11 & 1.16 & 474 \\
\hline
\end{tabular}
\caption{The results for $\sigma_{\gamma p \to J/\psi p} (W)$ in the
CDP, compare (\ref{14}). Explicitly, the results are based on
(\ref{15}) to (\ref{19}).}
\end{table}

Concerning the magnitude of $F_B(W^2)$ in (\ref{15}), we note that the
(logarithmic) variation of $\sigma^{(\infty)} (W^2)$ with energy in the
CDP is determined \cite{PRD85} by consistency of the $Q^2 \to 0$ limit
of the total cross section $\sigma_{\gamma^*p} (W^2,Q^2)$ with the 
experimental results on photoproduction, $\sigma_{\gamma p} (W^2)$.
The photoproduction limit in the CDP is given by \cite{PRD85}
\be
\sigma_{\gamma p} (W^2) = \frac{\alpha R_{e^+e^-}}{3 \pi} \sigma^{(\infty)}
(W^2) \ln \frac{\Lambda^2_{sat} (W^2)}{m^2_0},
\label{18}
\ee
where $m^2_0 = 0.15 GeV^2$. For $\sigma_{\gamma p} (W^2) \sim (\ln W^2)^2$,
the dipole cross section $\sigma^{(\infty)} (W^2)$ increases as a single
power of $\ln W$. According to (\ref{18}) for $F_B (W^2)$ in (\ref{15}),
we have
\be
F_B (W^2) \equiv \left( \frac{\sigma^{(\infty)} (W^2)}{\sigma^{(\infty)}
(W^2_1)} \right)^2 = \left( \frac{\ln \frac{\Lambda^2_{sat} (W^2_1)}{m^2_0}}
{\ln \frac{\Lambda^2_{sat} (W^2)}{m^2_0}} \right)^2 \left(
\frac{\sigma_{\gamma p} (W^2)}{\sigma_{\gamma p} (W^2_1)} \right)^2
\label{19}
\ee
We adopt a simple asymptotic Pomeron dependence for photoproduction,
$\sigma_{\gamma p} \simeq (W^2)^{0.08}$ \cite{Donnachie}, and from
(\ref{19}), we obtain the results for $F_B (W^2)$ given in the fifth
column of Table 1.

Our results for the energy dependence of the $J/\psi$ photoproduction
cross section following from (\ref{14}), according to (\ref{15}) and
(\ref{19}), are presented in the last column of Table 1. A comparison
to the experimental data in fig. 3 shows agreement of our results from
the CDP in Table 1 with the experimental results \cite{LHCb} from the
LHCb collaboration. 
The failure of the power-law fit in fig. 1 at large values of
the energy $W$ is an obvious consequence of the fact that in the CDP, due to
the denominator in (\ref{14}), the cross section is {\it not} proportional
to the square of the saturation scale, $(\Lambda^2_{sat} (W^2))^2$. This
behaviour may be traced back to the different dependence on the dipole
cross section in (\ref{1}) and (\ref{2}), respectively. 
For
a discussion on the connection between $\Lambda^2_{sat} (W^2)$ and the
gluon distribution in connection with $J/\psi$ production, compare ref.
\cite{PLB638}. As pointed out \cite{PLB638}, the $W$ dependence of 
$\Lambda^2_{sat} (W^2) \sim (Q^2/x)^{C_2}$ measured in $Q^2 = 0$
photoproduction (nevertheless) determines the $x$ dependence of the
gluon distribution function (gluon PDF) at sufficiently large values 
of $Q^2 \gsim M^2_{J/\psi}$, even though the simple proportionality of
the cross section to $\Lambda^4 (W^2)$ and the square of the gluon
distribution function does {\it not} hold in general.

In order to substantiate the interpretation of the experimental data,
it will be useful to return from photoproduction to the general case
of $Q^2 \ge 0$ in (\ref{14}). According to (\ref{14}), concentrating on
the $W^2$ dependence and suppressing a constant factor, we have
\bqa
&& \frac{d \sigma_{\gamma^* p \to J/\psi p} (W^2,Q^2)}{dt} \bigg|_{t \cong 0}
\propto \frac{(\sigma^{(\infty)} (W^2))^2}{\left(1 + \frac{Q^2+M^2_{J/\psi}}
{\Lambda^2_{sat} (W^2)}\right)^2}~~~ 
\frac{\Delta F^2(m^2_c, \Delta M^2_{J/\psi})}
{(Q^2 + M^2_{J/\psi})} \label{20} \\
&& \cong \left\{ \matrix{
\frac{\Lambda^4_{sat} (W^2) (\sigma^{(\infty)} (W^2))^2}{(Q^2 +
  M^2_{J/\psi})^2}~~~
\frac{\Delta F^2(m^2_c,~ \Delta M^2_{j/\psi})}{(Q^2 + M^2_{J/\psi})},
~~~~~~~~{\rm for}~~~~~~~~
\frac{Q^2 + M^2_{J/\psi}}{\Lambda^2_{sat} (W^2)} \gg 1, \cr
\hspace*{-1cm}
(\sigma^{(\infty)} (W^2))^2~~~ \frac{\Delta F^2 (m^2_c, \Delta M^2_{J/\psi})}
{(Q^2 + M^2_{J/\psi})},
~~~~~~~~{\rm for}~~~~~~~~
\frac{Q^2 + M^2_{J/\psi}}{\Lambda^2_{sat}
(W^2)} \ll 1.}\right.  \nonumber
\eqa
The $W$ dependence, according to (\ref{20}), at any fixed value of $Q^2 \ge 0$,
is determined by either $\eta_{c \bar c} (W^2,Q^2) \gg 1$ or
$\eta_{c \bar c} (W^2,Q^2) \ll 1$, respectively, where
\be
\eta_{c \bar c} (W^2,Q^2) \equiv \frac{Q^2 + M^2_{J/\psi}}{\Lambda^2_{sat}
(W^2)}
\label{21}
\ee
generalizes the low-x scaling variable \cite{PRD85} of deep inelastic
scattering (DIS),
\be
\eta (W^2,Q^2) \equiv \frac{Q^2 + m^2_0}{\Lambda^2_{sat} (W^2)}
\label{22}
\ee
to the case of heavy (charm) quarks. According to (\ref{20}), in analogy
to the total photoabsorption cross section, compare fig. 4, we have
either color transparency associated with $\eta_{c \bar c} (W^2,Q^2) \gg
1$ and $\sigma_{\gamma^*p \to J/\psi p} (W^2,Q^2) \sim \Lambda^4_{sat} (W^2)$,
or else we have saturation, $\eta_{c \bar c} (W^2,Q^2) \ll 1$ and
$\sigma_{\gamma^*p \to J/\psi p} \sim (\sigma^{(\infty)} (W^2))^2 \sim 
(\ln W^2)^2$.
The conditions $\eta (W^2,Q^2) \gg 1$ and $\eta (W^2,Q^2) \ll 1$
for the case of the total photoabsorption cross section in fig. 4,
translate into $\eta_{c \bar c} (W^2,Q^2) \gg 1$ and $\eta_{c \bar c}
(W^2,Q^2) \ll 1$, respectively.
\begin{figure}[h]
\begin{center}
\epsfig{file=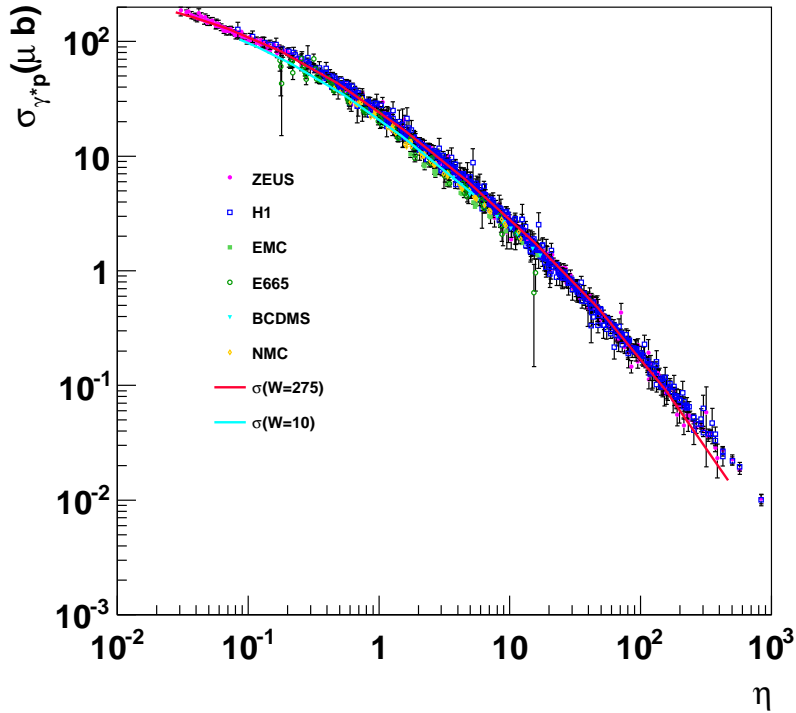,width=12cm}
\vspace*{-1cm}
\caption{The total photoabsorption cross section as a function of
\hfill\break
$\eta (W^2,Q^2) = (Q^2 + m^2_0)/\Lambda^2_{sat} (W^2)$ (from ref. 4).}
\end{center}
\end{figure}

We return to $Q^2 = 0$ photoproduction. In Table 1, third column, we see that
for the energy range of the experimental data, the variable $\eta_{c \bar c}
(W^2, Q^2 = 0) = M^2_{J/\psi}/\Lambda^2_{sat} (W^2)$ covers the range of
\be
2.2 \gsim \eta_{c \bar c} (W^2,Q^2) \ge 0.43.
\label{23}
\ee
This range corresponds to a region in the vicinity of $\eta_{c \bar c}
(W^2,Q^2) = 1$ that determines the transition from color transparency
to saturation.\footnote{Our analysis leading to Table 1 was based on (\ref{14})
or the first line in (\ref{20}), or equivalently (\ref{15}). The
approximations in the second line of (\ref{20}) in this connection must
be considered as qualitative indications. They need a larger range in $Q^2$
and $W^2$ to become quantitatively reliable.} The (approximate)
$\Lambda^4_{sat} (W^2)$ power law dependence, with increasing energy, breaks
down to turn into a weak $(\ln W^2)^2$ dependence.

The deviation of the LHCb experimental data in fig. 3
from the H1 power-law fit
at $W \gsim 500 GeV$ provides empirical evidence for the transition
from color transparency to saturation. An extension \cite{Jones} of a
simple proportionality, $\Lambda^4_{sat} (W^2) \propto \left(\alpha_s(Q^2) x
g (x, Q^2)\right)^2$, of the cross section of $J/\psi$ production, 
$\sigma_{\gamma^*p \to J/\psi p} (W^2,Q^2)$, at
fixed $Q^2 \ge 0$ to large values of $W$ does {\it not} lead 
to a decent high-energy
behaviour, such as $(\ln W^2)^2$, of the cross section 
in the true limit of $W \to \infty$ that 
is required to hold, however, and in fact is guaranteed by 
saturation as contained
in the color-dipole picture.





\end{document}